# Flexible Amorphous Superconducting Materials and Quantum Devices with Unexpected Tunability

Mohammad Suleiman,[1,2] Emanuele G. Dalla Torre[3] and Yachin Ivry[1,2,*]

[1] Department of Materials Science and Engineering, Technion – Israel Institute of Technology, Haifa 3200003, Israel.

[2] Solid State Institute, Technion – Israel Institute of Technology, Haifa 3200003, Israel.

[3] Department of Physics, Bar-Ilan University, Ramat Gan 5290002, Israel

*Correspondence to: ivry@technion.ac.il.

In superconductors, electrons exhibit a unique macroscopic quantum behavior, which is the key for many modern quantum technologies. Superconductivity stems from coupling between electrons and synchronized atom motion in the material. Hence, the inter-atomic distance and material geometry are expected to affect fundamental superconductive characteristics. These parameters are tunable with strain, but strain application is hindered by the rigidity of superconductors, which in turn increases at device-relevant temperatures. Here, we developed flexible, foldable and transferable superconducting materials and functional quantum nanostructures. These materials were obtained by depositing superconductive amorphous-alloy films on a flexible adhesive tape. Specifically, we fabricated flexible superconducting films, nanowires and quantum interference devices (SQUIDs) and characterized them under variable magnetic-field, current, temperature and flexure conditions. The SQUID interference periodicity, which represents a single flux quantum, exhibits unprecedented and unexpected tunability with folding curvature. This tunability raises a need for a relook at the fundamentals of superconductivity, mainly with respect to effects of geometry, magnetic field inhomogeneity and strain. Our work supplies a novel platform for quantum and magnetic devices with local tunability.

**Introduction**

Developing flexible electronic devices is a recent effort (*1–3*) that involves a 'win-win situation,' as follows. Materials whose functionality is unaffected by the bending lead to devices that operate robustly under variable strain conditions. Alternatively, strain is used as a tuning parameter for functional properties. Yet, most electronic devices are currently made of rigid inorganic materials that undergo irreversible deformation under flexure. Moreover, because material rigidity increases with decreasing temperature, developing flexible inorganic quantum materials that operate at low temperatures, such as superconductors, is a major challenge. In superconductors, the electric properties are intertwined with collective mechanical vibrations of the atoms. Thus, not only does the inability to strain superconductors and change the inter-atomic distance encumber technological developments, but it also prevents us from examining fundamental behavior of superconductivity. Furthermore, there is the strong dependence of superconductivity on geometry, *e.g.*, in magic-angle superconductivity in bilayer graphene (*4*, *5*), as well as in miniaturized structures (*6–8*). Hence, geometrical tunability by means of flexure is also likely to affect fundamental superconductive properties.

The current race towards superior quantum sensing, communication and computation technologies hinges strongly on compatible miniaturized superconductive devices (*9–12*). The functionality of superconductive materials stems from their unique magnetic properties, lack of electric resistance and macroscopic quantum behavior that emerges at an abrupt transition. Recently, several significant milestones have been reached owing to these properties, while they are all based on superconductive technologies. Examples include quantum supremacy in data processing (*13*), qubits with 100-μsec decohrence time (*14*, *15*), ultra-sensitive nanoscale magnetic metrology (*16*), and superior single-photon detectors that allow long-distance quantum key distribution (*17–21*), and quantum-optics data manipulation (*22–25*) .

Despite these accomplishments, tuning superconducting characteristics, especially locally at a single-device lengthscale, is a hurdle for several reasons. First, although the absence of resistance in superconductors is beneficial for some applications, the inability to apply local gate voltage (as in the case of semiconducting transistors) forces us to replace electric-field biasing with magnetic and microwave operation, which is large scale and hence non selective. Other challenges include limited scalability, versatility, and robustness of materials and device fabrication, as well as data transfer between human-compatible ambient and the low device-operation temperature. Hence, there is a need for novel superconducting-based quantum platforms that overcome these challenges, mainly by enabling, at last, local tunability of the quantum and functional properties of superconductors. An ideal platform would therefore allow operation above liquid-helium temperature, easy integration with other circuit elements, simple device fabrication, mechanical and electro-magnetic robustness, as well as non-magnetic tunability of individual devices, *e.g.* by means of strain. Moreover, because the behavior of a flexed superconductor is yet unknown, such a platform is useful for characterizing fundamental properties of superconductors that are difficult or impossible to realize otherwise.

Here, we demonstrate a novel platform for quantum superconductive devices by using the relatively low synthesis temperature and good adhesion of amorphous-alloy superconductive films (*26*). Metallic-glass superconductors were processed on flexible substrates. The new structure revealed unexpected tunability of SQUID oscillations with mechanical bending. We found that the interference periodicity decreases with increasing flexure curvature, giving rise to > 12-fold periodicity difference between bent and flat SQUIDs. We demonstrate reproducible tunability for different materials and device geometries, while characterizations of complimentary superconducting properties show no significant dependence on the bending that supports the order of magnitude periodicity enhancement. Using the adhesive nature of the flexible substrate, we show that the devices are transferrable without any noticeable change in their performance. Robust device operation is shown for temperatures ranging from $T$=20 mK to $T$ > 5.7 K, $B$ > 6 T magnetic fields and 4 to 25 nm film thickness. Finally, we show substrate insensitivity of the amorphous SQUIDs and propose additional paths for individual device tunability by means of external electric field with ferroelectric-piezoelectric substrates.

**Background**

Prior to delving into the experimental system, we should first introduce how quantum properties of superconductive structures are measurable and which of them are tunable in potential. The basic building block of a superconductive quantum-device is a Josephson junction or a weak link. In such a structure, superconductivity is absent or suppressed. Superconducting currents flow across a weak link by the tunneling of electron Cooper-pairs between the electrodes that sandwich the junction. A quantum state of a junction is characterized by a complex order parameter $\psi = \Delta e^{i\theta}$. Here, the amplitude ($\Delta$) sets the maximal superconducting current and the critical temperature ($T_c$) of the junction. Therefore, is measurable with a single junction. Contrariwise, the phase (θ) cannot be measured directly in absolute values, but only phase differences are measurable. Typical quantum superconductive devices comprise two weak links that are connected in parallel. The phase difference between them is periodic with external magnetic fields, resulting in an interference pattern, which is reminiscence of the light interference in a two-slit experiment. The interference pattern in such superconducting quantum interference devices (SQUIDs) is obtained when the current at which the material switches from superconductor to normal metal ($I_s$) oscillates as a function of applied magnetic-flux density $B_{ext} = \Phi_{ext}/A$. Here, $A$ is the device area and $\Phi_{ext}$ is the resulting flux that is perpendicular to the SQUID. The periodicity of this oscillation ($B_0$) is constant and corresponds to a single flux quantum ($\Phi_0 \equiv h/2e$, $h$ is Planck's constant and $e$ is the electron charge), while satisfying:

$$\Phi_0 = B_0 A \quad (1)$$

SQUIDs are used for magnetic sensors with high accuracy that spans a broad range of magnetic fields, where $B_0$ dictates the magnetic-field sensitivity of the device. Asymmetric SQUID structures that often comprise capacitors and inductors demonstrate high-frequency resonance, which is the basis for superconducting quantum and low-power data processing. The resonance frequency is set by the constant periodicity ($\Phi_0$) of the dc SQUID. Tuning the functional properties of a SQUID requires therefore adjustability of Δ, θ or $\Phi_0$. In SQUIDs, among these three parameters, θ is the easiest to modulate because it is determined by the external magnetic flux. To minimize the areal footprint, magnetic fields are often induced by a current flowing in a nearby wire (*27*), though such devices are still orders of magnitude larger than electric-field gating in semiconductors. Nevertheless, to-date, there are only limited methods to tune Δ (*28*, *29*), while there are no available methods to tune $\Phi_0$ for a given device. Thus, examining the effects of geometrical variation on these parameters, *e.g.* by flexure, is an important task. Likewise, within the present work, we used the high sensitivity of SQUIDs to magnetic fields for characterizing fundamental properties of flexed superconductors that are difficult or impossible to realize otherwise.

**Experimental**

Amorphous molybdenum silicide and tungsten silicide films (4-1000 nm in thickness) with respective $\alpha$Mo$_{81}$Si$_{19}$ and $\alpha$W$_{60}$Si$_{40}$ stoichiometry were deposited on a flexible polymer substrate by means of magnetron sputtering, while the former was also sputtered on silicon-chip, glass, ferroelectric and layered-material substrates (see Materials and Methods for details). The amorphous structure made the requirement of substrate-film lattice matching that is typical for crystal unnecessary and helped obtain good adhesion of the films to the different substrates. Likewise, the relatively low growth temperature helped protect the substrate, *e.g.* in the case of the polymer. Film thickness, amorphous structure and chemical composition were determined with ellipsometry, X-ray diffraction (XRD) and X-ray photoelectron spectroscopy (XPS, for molybdenum silicide films), respectively (Figures S1-2). Planar SQUIDs were fabricated with electron-beam lithography with high yield on various substrates. Samples were characterized electrically in three different systems with high reproducibility (see Materials and Methods).

We used the reduced film thickness of the amorphous superconductors for obtaining geometrical-assisted flexibility (*30*). To allow in-situ characterization of bent devices, samples were synthesized on flexible polyamide adhesive tapes with a thickness $t=30$ μm. Such tapes maintain their mechanical elasticity and adhesion at the superconducting-relevant low temperatures, while they also remain thermally and electrically insulating. To demonstrate the effect of flexure, devices were placed on holders with various radii of curvature ($r$) while samples were transferred between the holders using the adhesive nature of the flexible tapes. The SQUIDs were placed with the weak links parallel to the arcing cylinder circumference (Figure 1), allowing direct characterization of devices under variable strain conditions.

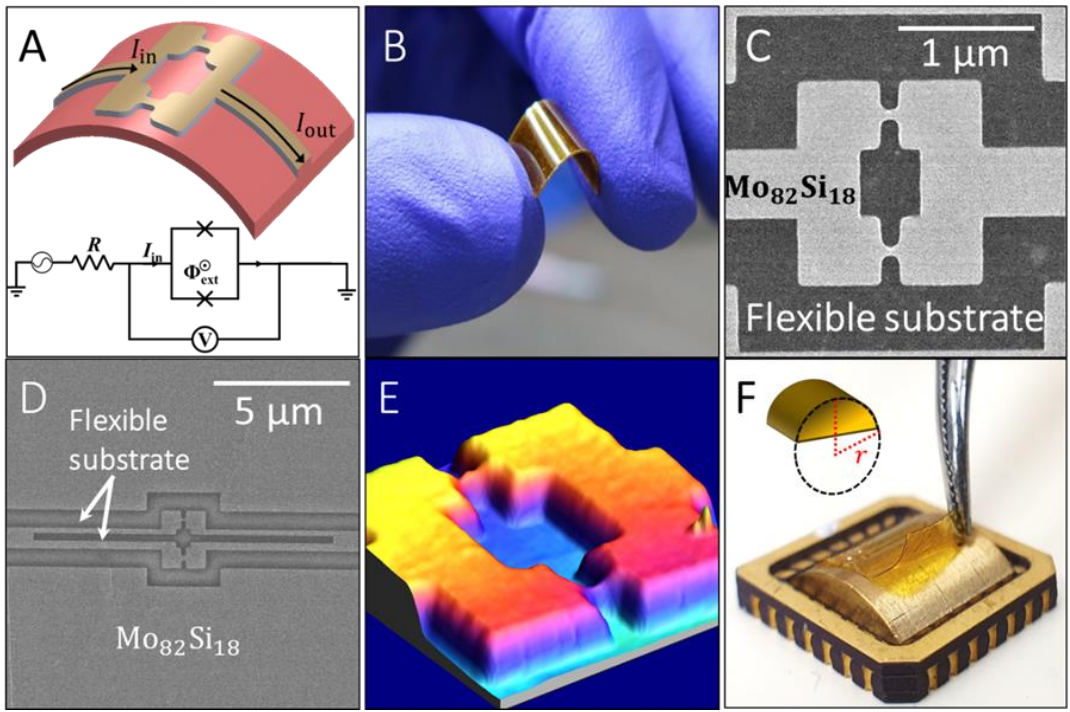

**Figure 1| Flexible superconductive devices and films.** (**A**) Schematics of a flexible superconducting quantum interference device with current circulating in a loop that contains two parallel weak links. (**B**) Optical photo of foldable superconductive devices on a flexible polyamide tape. (**C**) Electron micrographs of 15-nm thick planar $\alpha$Mo$_{81}$Si$_{19}$ SQUIDs with small square and (**D**) large active areas on a flexible polyamide substrate. (**E**) Atomic force microscopy profiling of a 23-nm thick amorphous-alloy SQUID (on a silicon chip), showing

the device topography (1.5 × 1.5 μm² scan area). (**F**) Optical photos of an amorphous-alloy SQUID on flexible and transferrable adhesive polyamide tape that is placed on a sample holder with $r = 11.2$ mm (insert: schematics of the radius of curvature of the bent holders).

**Main Results: effects of flexure on SQUID behavior**

The first milestone on the way to a flexible SQUID was obtained by demonstrating that amorphous-alloy superconductors can be used for SQUIDs. Figure 2 shows the effect of bending on flexible SQUIDs. Figure 2A demonstrates the interference pattern of a flat $\alpha$Mo$_{81}$Si$_{19}$ SQUID with 20.42 mT periodicity (the device geometry follows Figure 1B). Here, the switching current is of the order of 3-4 μA, which is significantly smaller than crystalline devices of similar geometries (*25*, *28*, *29*). Bending the same device by transferring the adhesive tape to sample holders of different curvatures and placing the SQUID at the center of the holder changed the interfering periodicity. Figure 2A shows that bending the sample with $r$=17.5 mm and $r$≈1.5 mm reduced the interference periodicity to 16.06 mT and 1.44 mT, respectively, indicating on an order of magnitude change between curved and flat devices. The dependence of periodicity on bending curvature ($\kappa \approx 1/r$) is summarized in Figure 2B. This graph shows linear reduction in periodicity for small values of $\kappa$.

To confirm that the change in periodicity is universal and not due to a specific device geometry or material, Figure 2F contains also data from a SQUID of a different geometry with a larger loop area (following Figure 1D). Likewise, Figure 2B shows a similar behavior for a flexible $\alpha$W$_{60}$Si$_{40}$ SQUID (square geometry). Interference measurements for additional curvatures are given in Figure S3.

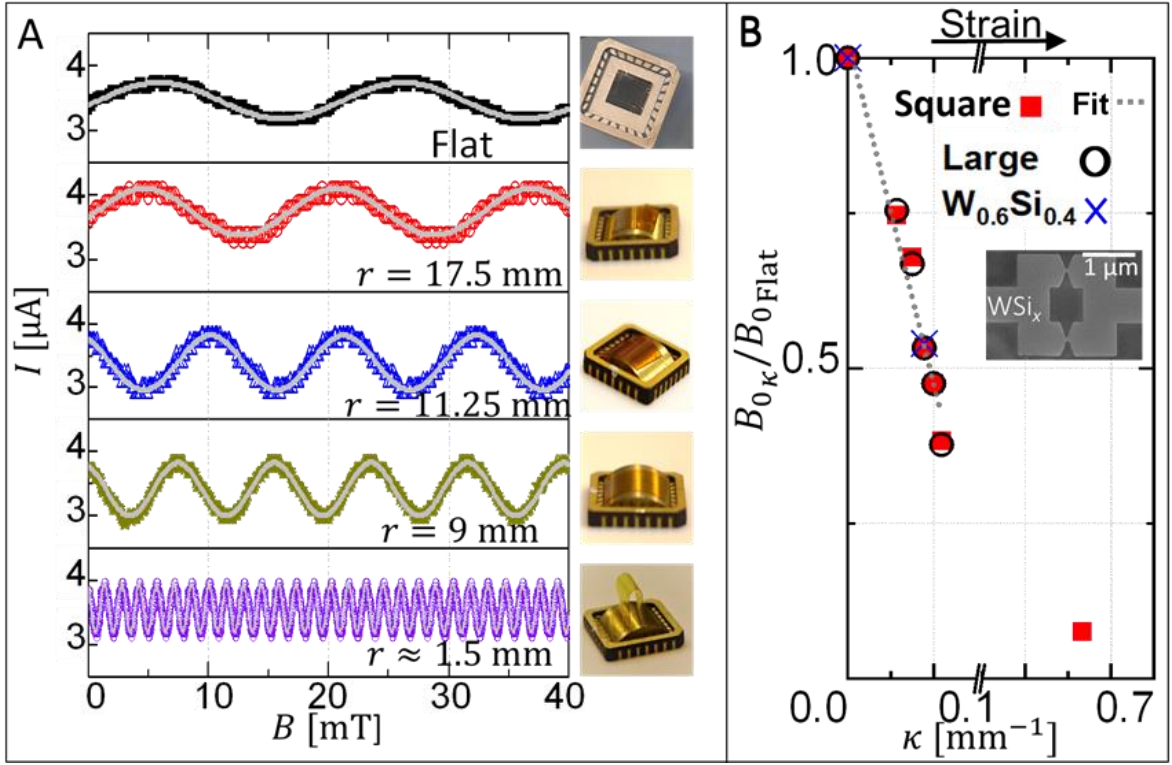

**Figure 2| Effects of bending on interference periodicity in flexible SQUIDs**. (**A**) Interference profile of SQUIDs with increasing curvature (top-to-bottom) at 3 K. The flat device exhibits 20.42 mT periodicity. Bending the device with $r$=17.5 mm, $r$=11.25 mm and $r$=9 mm, reduces the periodicity to 16.06 mT, 11.06 mT and 8.02 mT, respectively. Further folding to $r$ =1.5 mm reduces the periodicity to 1.44 mT. Best fits to a sine function are given as solid lines for each interference pattern (see Figure S3 and Table S1 for details). Optical photos of the sample at the corresponding bending states are also given. (**B**) Periodicity (normalized) as a function of curvature for devices of small and large geometries (following Figure 1C and 1D, respectively) as well as of a SQUID from a different material ($\alpha W_{60}Si_{40}$), showing the universality of the behavior (see Table S2). Here, the periodicity decreases linearly with curvature for large $r$ values. Overlaid dashed line is best fit: $B_{0_\kappa}/B_{0_{Flat}} = 1.007 - 7.05\kappa$, where $B_{0_\kappa}$ and $B_{0_{Flat}}$ are the periodicities of the curved and flat devices, respectively, and $\kappa$ is given in mm$^{-1}$ (insert: $\alpha W_{60}Si_{40}$ SQUID on polyamide).

**Complementary results: effects of flexure on superconductive properties**

To examine the effect of bending on superconductive properties, we processed 2000×100×15 nm$^3$ nanowires that are physically placed near the SQUIDs and tested them under the same flexing conditions of Figure 2A (wires were parallel to the bending direction). We current-biased the wires and measured the resultant voltage under variable temperature and magnetic field conditions. Figures 3A-C show the superconductive critical values of magnetic-field vs. current, and temperature, as well as critical current vs. temperature, for various bending conditions. Figure 3D summarizes the three-dimensional superconductive magnetic-field – current – temperature surface for different bending radii, while the cooling curves of the flexed devices are given in Figure 3E. Representative current-voltage data are presented in Figure S4. Hall measurements of continuous films were performed for various curvatures (Figure 3F), while the extracted critical magnetic field, current and temperature as well as coherence length, and electron density are given in Table 1.

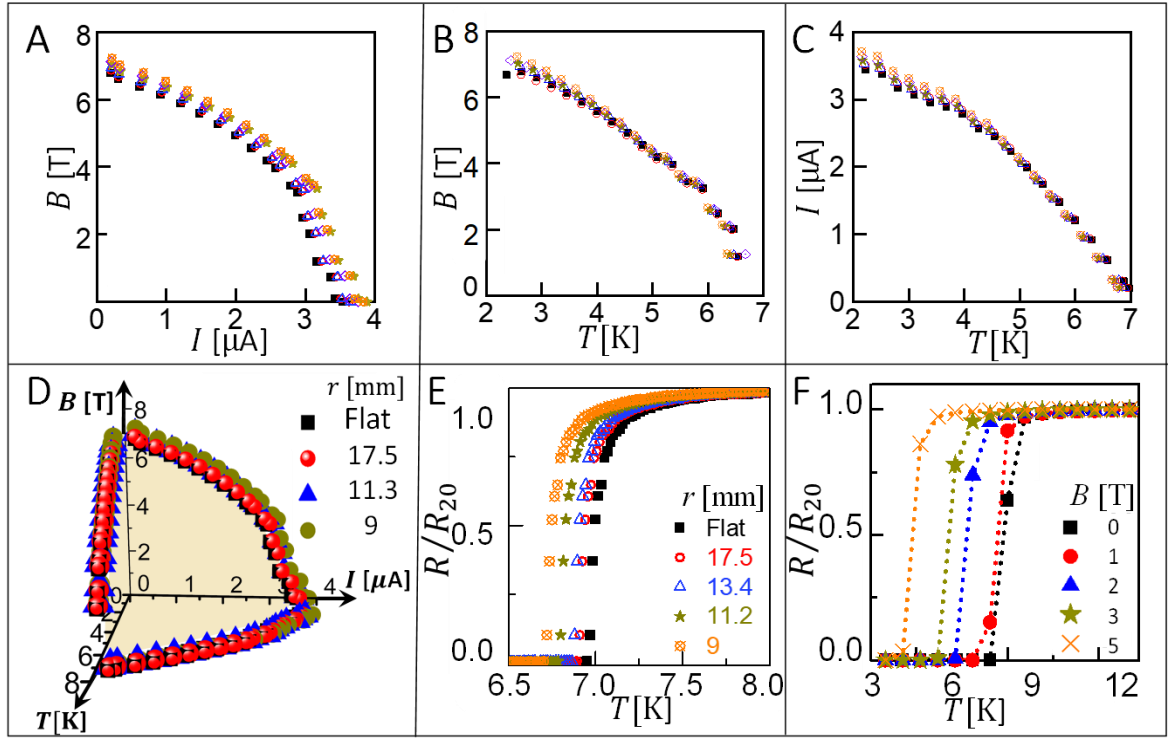

**Figure 3|Effects of flexure on superconducting properties**. (**A**) Critical magnetic field vs. current (at 3 K) and (**B**) vs. temperature, as well as (**C**) critical current vs. temperature in 2000×100×15 nm$^3$ $\alpha$Mo$_{81}$Si$_{19}$ wire for various curvature conditions. (**D**) Superconductive critical surface and (**E**) cooling curves of the same wire for representative curvature values. Legend represents color scheme also in (A-C). (**F**) Hall resistance of a flexible superconducting film for various values of $B$ (data from other flexure conditions are presented in Table 1).

| $r$ [mm] | $\kappa$ [mm$^{-1}$] | $T_c$ [K] | $I_c$ [μA] | $B_c$ [T] | $\xi$ [nm] | $n_e$ [$10^{29}$m$^{-3}$] | $2\Delta$ [meV] |
|---|---|---|---|---|---|---|---|
| ∞ | 0 | 6.97 | 3.55 | 7.77 | 6.51 | 3.24 | 2.12 |
| 17.50 | 0.06 | 6.91 | 3.63 | 7.95 | 6.44 | 3.32 | 2.10 |
| 13.33 | 0.08 | 6.88 | 3.65 | 8.02 | 6.41 | 3.37 | 2.09 |
| 11.25 | 0.09 | 6.8 | 3.89 | 8.13 | 6.37 | 3.58 | 2.07 |
| 10.00 | 0.10 | 6.73 | 3.79 | 8.21 | 6.33 | 3.49 | 2.05 |
| 9.17 | 0.11 | 6.62 | 3.84 | 8.38 | 6.27 | 3.54 | 2.01 |

**Table 1| Effects of curvature on superconductive properties.** Critical temperature, current and magnetic field as well as coherence length ($\xi$) of 2000×100×15 nm$^3$ wires as a function of curvature. These parameters are calculated for $T$=0 K from the data presented in Figures 3A-E. The film electron density at the normal state ($n_e$) and Cooper-pair binding energy (2Δ) that were extracted from Hall measurements (Figure 3F) are also presented.

Figures 3A-B show that the critical magnetic field ($B_C$) increases with increasing $r$ for all temperatures, while we extracted that at $T$=0 K, $B_c(r=\infty)$=7.77 T and $B_c(r$=9 mm)=8.4 T (see Table 1). Likewise, Figure 3D shows that at 1.9 K, $I_s(r=\infty)$=3.55 µA and $I_s(r$=9 mm)=3.84 µA. As opposed to the trend in $B_c$, the difference in $I_s$ between small and large curvatures reduces with increasing temperature until there is a crossover at $T$=5.4 K, above which devices with smaller curvature have a lower $I_s$ than with higher curvature.

**Discussion**

Despite the strong effect of bending on the SQUID interference pattern (> 1200%, Figure 2), changes in $I_s$, $B_c$ and $T_c$ in these measurements are no more than 10% with respect to the flat-wire (Figure 3). Thus, it is difficult to provide a simple explanation for the vast change in periodicity based on these parameters.

Following Equation 1, the interference periodicity of a given device varies only with the device area, $A$, and the change is linear. However, in the current experiments, $A$ is effectively constant. That is, the reduction in $A$ due to bending is negligible because the SQUID length is small (only a few hundreds of nanometers) in comparison to the large radius of curvature (a few millimeters). In addition, reduction in device periodicity (*i.e.*, smaller $B_0$) corresponds to an increase in $A$. In contrast, bending a SQUID reduces the effective device area that is perpendicular to $\Phi_{ext}$, if at all. Moreover, the geometry insensitivity of the change in $B_0$ that was presented in Figure 2E for small square and large rectangular devices helps rule out the changing-area hypothesis.

Presumably, the device area may increase due to strain. However, the strain in the examined geometry is: $\epsilon = t/r \leq 0.02$. This strain is also too small to allow meaningful geometric deformations that will give rise to periodicity reduction in the framework of Equation 1. Yet, this strain value may be meaningful in other manners, *e.g.*, by affecting the tunneling at the weak link, especially when bearing in mind that there may be high strain concentration at the weak link. Nevertheless, exact effects of high strain concentrations on weak links and SQUIDs, as well as on general superconducting materials are still unclear.

Equation 1 allows an additional scenario for decrease in periodicity—if the perpendicular magnetic field varies between the different experiments. That is, when either the magnitude or direction of the magnetic field varies as a result of the bending. We ruled out this hypothesis experimentally in several ways. Figure S5 shows reproducible measurements from three different testing systems. Next, we negated contribution of the cylindrical-holder material as well as of the effects of height of the holder on *e.g.*, magnetic-field inhomogeneity. That is, the SQUID and superconductive properties were measured when the device was placed on a flat holder of the same brass material, which raised the sample in 2 mm with respect to the flat device. Figure S6 shows that the resultant interference of elevated devices was identical to the flat sample that was measured without the holder, and not to a bent device.

Magnetic-field variation may arise also due to lensing effects as a result of current flow in the curved superconductor. Following the magnetic-lensing simulations of Prigozhin *et al.* (*31*), we found that the maximal possible contribution of such lensing effect in our devices is smaller than 5%, and hence cannot explain the above observations (Figure S7). Likewise, the magnetic screening of the superconducting film in the area outside the device may contribute to the enhancement. However, this contribution is expected to be proportional to the change in the effective area that is perpendicular to the magnetic field. The maximum difference in this area may be from a 5-mm chip size in the flat device to roughly 2-3 mm in the highly-curved device ($r$=1.6 mm), which is still not close to the order of magnitude enhancement in interference periodicity. Moreover, if the current that flows in the wires or the magnetic field in the ambient enhance the magnetic field within the SQUID due to bending, then the interference patterns in Figure 2A may demonstrate strong non-linearity as a function of $B_{\text{ext}}$, either in periodicity or in $I_s$.

To ensure the absence of artefacts by an independent measurement we rotated the devices in 90° on a bent sample holder. Here, the weak links were unstrained, aligning perpendicularly to the bending (*i.e.*, to the cylindrical circumference of the holder) as illustrated in Figure S8A. Figure S8 shows clearly that the measured interference pattern of the rotated device was identical to the flat SQUID, which in turn is very different from the interference profile when the SQUID was placed on the same holder, but with the weak links aligned along the straining direction (Figure 2). We should note that recent studies (*32*) suggest that critical current in $Mo_{80}Si_{20}$ wires is determined by two competing mechanisms of vortex crossing and Cooper-pair breaking, while the dominancy of these mechanisms is affected by external magnetic fields. However, although these mechanisms affect *e.g.*, the performance of superconductive nanowire single-photon detectors (*33*), presumably, Equation 1 is indifferent to these mechanisms.

In addition to unknown effects of strain on weak links, a possible cause for periodicity change that we cannot eliminate at this point is the effect of magnetic-field inhomogeneity on the superconductor. That is, although in our systems the magnetic field is rather homogeneous in magnitude, it is inhomogeneous with respect to the direction. Such inhomogeneities in thin conducting films are predicted to confine the charge carriers, as well as induce weak localization, interference effects, universal conductance fluctuations and local conductance variations (*34–36*), but their effects on superconductors are not yet known, especially for bent geometries. From the perspective of the film outside the device, the magnetic field direction changes substantially (changes at the device lengthscale are negligible). For instance, for $r$=1.5 mm (Figure 2), the direction is parallel at the areas of the tape that stand vertically, changes gradually to perpendicular at the center of the curved area (near the SQUID) and then changes again to become parallel to the film. This inhomogeneity (which is different from a lensing effect due to current flow in a curved superconductor) may be accompanied by a significant enhancement in current concentration near the SQUID, and hence responsible for the enhancement in effective magnetic field that the SQUID senses. These effects comply with our simulations (Figure S7) of ~400% magnetic-field enhancement at the edge of a thin-film geometry with respect to the body of the film. However, this enhancement by itself does not change with curvature. Hence, understanding the exact effects of magnetic-field inhomogeneity requires more dedicated future attention.

**Conclusions**

Our results show that superconducting films and nanostructures operate under varying flexure conditions, while the flexure introduces strong unpredicted magnetic enhancement over a broad range of thicknesses and temperatures (Figure S5), as well as magnetic fields (Figure S10). The results also show that SQUIDs can be made of metallic-glass superconductors and that these devices demonstrate robust properties and processing conditions, while the reduced switching currents (~3 μA) are useful from an application perspective.

The mechanism in which bending strain and curved surfaces affects the amorphous superconductive devices and gives rise to significant variation in SQUID periodicity is currently unclear. Yet, the mechanical tunability of the device properties along with substrate insensitivity (Figure 4A) expands the realm of superconductive quantum technologies and magnetic devices. Moreover, the mechanical tunability can assist upscaling of existing technologies. Ideally, devices should be strained individually (Figure 4B). Figure 4C shows 15-nm thick amorphous $\alpha$Mo$_{81}$Si$_{19}$ coating of a 2-inch substrate. Insignificant changes in properties were measured for devices that were fabricated at different areas, demonstrating potential for upscaling. Similarly, we coated a highly tetragonal (001) 40-nm thick PbZr$_{0.1}$Ti$_{0.9}$O$_3$ (PZT) ferroelectric film (on a DyScO$_3$ substrate, with SrRuO$_3$ bottom electrode) that is known for its large remnant polarization and high piezoelectric coefficient (*37*). Figure 4D shows that in these materials, the $T_c$ (and hence Δ) is tunable with external electric field. Previous studies proposed that in superconducting-ferroelectric stacks, ferroelectricity affects the superconductive properties by the surface charge that the ferroelectric polarization domains form (*38–42*). However, given the above observations, mechanical deformation of the piezoelectric material may also affect $T_c$. Therefore, we propose that for future applications, individual SQUIDs will be tunable electrically, using micro electro-mechanical systems (MEMS), including piezoelectrics (Figure 4). The convenient growth conditions of amorphous superconductors assist maintaining a clean interface with the ferroelectric, which is important for such devices.

To further demonstrate the potential of the system that is presented in Figure 4B, we examined the repeatability of the superconductive device behavior. Figure S9 shows the interference pattern of an $\alpha$Mo$_{81}$Si$_{19}$ SQUID that is placed on a sample holder with $r$=11.2 mm. The sample was detached from the sample holder and re-attached again nine different times (each removal-attachment cycle included wire bonding), allowing us to examine the effects of bending and replacing on the superconductive device behavior. The interference pattern of the first placement is compared with the interference of the sixth and ninth placements, showing no changes between the graphs. Note that more than additional twenty removing-reattaching cycles were done between these nine cycled of this holder for measuring the device on sample holders with other curvatures.

Lastly, thanks to the unique mechanical and magnetic properties of flexible superconductors on adhesive tapes, these materials are promising for magnetic-field shielding, as well as for magnetic-field production (*i.e.*, superconducting coils), mainly for geometries that may be complicated to comply with otherwise.

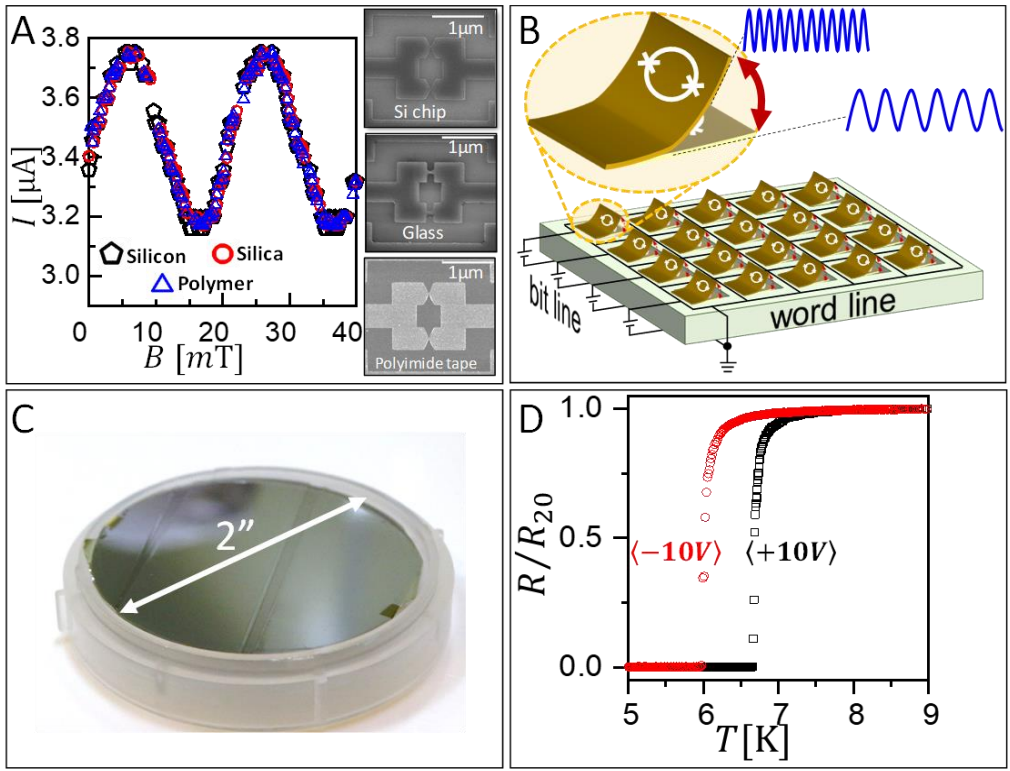

**Figure 4|Proposed integrated superconductive flexible devices**. (**A**) Interference pattern of flat $\alpha$Mo$_{81}$Si$_{19}$ SQUIDs on polyamide tape, fused silica and Si-chip substrates, showing device robustness and substrate insensitivity. (**B**) Proposed system with superconductive quantum devices that are controlled individually. (**C**) $\alpha$Mo$_{81}$Si$_{19}$ coating of polyamide tape that covers a 2-inch wafer. (**D**) Cooling curve of $\alpha$Mo$_{81}$Si$_{19}$ on a ferroelectric voltage shows changes in $T_c$ with external electric field.

**Acknowledgments**

The authors would like to acknowledge financial support from the Israel Science Foundation (ISF) grants number 1602/17 and 154/19, the Zuckerman STEM Leadership Program, the Technion Russell Barry Nanoscience Institute and the Technion Microelectronic Center. We thank the following personnel for technical support: Dr. Guy Ankonina (sputtering and ellipsometry); Dr. Adi Goldner (device fabrication); Dr. Kamira Weinfeld (XPS); Dr. Valentina Korchnoy (SEM) as well as Prof. Hadar Steinberg, Dr. Devidas Taget Raghavendran and Dr. Tom Dvir, Prof. Eyal Buks and Dr. Anna Eyal from the Technion's Quantum Materials Research Center (low-temperature measurements). Likewise, we thank Prof. Nava Setter and Dr. Barbara Fraygola for supplying us with the PZT films, as well as Dr. Hemaprabha Elangovan and Mr. Bader Zaroura for assisting with the graphical demonstrations and optical footage, respectively. Finally, we thank Prof. Karen Michaeli and Prof. Daniel Podolsky for fruitful discussions regarding theoretical aspects of the work as well as Prof. Moti Segev and Prof. Karl Berggren for helping sharpen the manuscript.

## Materials and Methods
### Film deposition

Superconducting films were deposited using ATC2200 (AJA International inc. MA, USA) off-axis magnetron sputtering. $Mo_{80}Si_{20}$ and $W_{60}Si_{40}$ targets (99.95%) were purchased from AJA International. Sputtering conditions included 35W (0.27 Å sec$^{-1}$ deposition rate), 3 mTorr and 50 sccm Ar flow (99.9999%), with 20-cm target-sample distance at 22° C. Film thickness was determined by using a five-oscillator fit (one Drude, one Tauc-Lorentz, one Gauss-Lorentz and two Gaussians) of varying-angle spectroscopy ellipsometry (VASE) measurements of the optical constants (M-2000 by J. A. Woollam, NE USA).

Substrates were purchased from commercial suppliers: intrinsic Si with 318 nm oxide layer (UniversityWafers Inc.); glass substrates (microscope slides by Corning® Inc., NY, USA); layered mica (Asylum Research by Oxford Instruments Ltd.); and polyimide (20-μm width adhesive polyamide tape from Zhuhai Store, Gaungdong, China). Thirty nine chips were produced (13 on silicon, 5 on glass, 14 on polyamide and two on mica for $\alpha Mo_{81}Si_{19}$ and 2 $\alpha W_{60}Si_{40}$ films on polyamide), with a typical size of 5 × 5 mm$^2$.

### Film characterization

XRD profiling was done using θ − 2θ with a Rigaku SmartLab 9-kW high-resolution diffractometer. A Cu kα rotating-anode source at 45 kV tube voltage was used, with a 150-mA tube current as well as with a 0D silicon drift detector.

Stoichiometry analysis of the molybdenum silicide was done with a UHV (2$10^{-10}$ Torr) XPS (Versaprobe III – PHI Instrument, PHI, USA). Samples were irradiated with a focused X-ray Al $K$α monochromated source (1486.6 eV, beam size 200 μm, with 25 W and 15 kV). Outcoming photoelectrons were directed to a spherical capacitor analyzer (SCA). Sample charging was compensated by a dual-beam charge neutralization based on a combination of a traditional electron flood gun and a low energy argon ion beam.

Topography characterization was done with tapping-mode atomic force microscopy (MFP3D Asylum Research by Oxford Instruments Ltd.), using a silicon tip with 70-kHz resonance frequency and 2 N m$^{-1}$ spring constant, while WSxM (*43*) was used for presenting the data.

$T_c$ and other critical values were determined by using the 90% drop from the normal resistance at 20 K, while the sheet resistance was measured with a 4-point probe station (Signatone Co., CA USA) connected to a Keithley 6220 multimeter (Keithley Instruments, Tektronix OH, USA). Typical $T_c$ was measured as 6.7-7.2 K and 4-4.5 K with corresponding ~500 Ω/□ and ~450 Ω/□ sheet resistance for 15-nm molybdenum silicide and tungsten silicide films, respectively.

### Device fabrication

Devices (SQUIDs and nanowires) were fabricated by the electron-beam lithography method. PMMA 950 A3 electron-beam resist was applied by spin coating, followed by 2-min post baking. A thin layer of '*e*-spacer' was used occasionally to avoid charging. Beam conditions were 610 μC cm$^{-2}$ dose of 250 pA at 100 kV (Raith EBPG 5200). After writing, PMMA was developed with MIBK, followed by topography profiling with Alpha-Step 500 (KLA-Tencor Inc.). Reactive-ion etching (RIE, Plasma Therm 790) with $CF_4$ gas was used for transferring the pattern to the layer.

Chips with films and devices were glued to a holder and wire-bonded with aluminum wires (VB16 wedge bonder, Micro Point Pro Ltd.).

**Cryogenic testing**

Cryogenic electric and magnetic measurements were done with three different systems: (i) DynaCool Quantum Design Inc. (CA, USA); (ii) BlueFors (Helsinky, Finland) with MFLI lock-in amplifier (Zurich Instruments, Switzerland); (iii) in-house made cryogenic system with Keithley 236 (Keithley Instruments, Tektronix OH, USA) and Yokogawa 651 (Yokogawa Electric, Tokyo, Japan) multimeters and an SR830 (Stanford Research Systems, CA, USA) lock-in amplifier. SQUID measurements were done at 100-12000 Hz, allowing examination of measurement reproducibility.

A typical chip contains 16 SQUIDs and 24 wires, while 3 chips contained large SQUIDs (on average, we characterized three SQUIDs for a chip (wires were measured in three chips, while we characterized three wires for each chip). More than 15 devices were measured for variable curvatures. Most devices were measured for ~three cycles of transferring the samples between the holders of different curvatures. All of these measurements were reproducible and demonstrated a similar behavior to the results that are discussed in this work.

More details regarding sample preparation and characterization can be found elsewhere (*28*, *29*, *44*).

**Supplementary Materials**

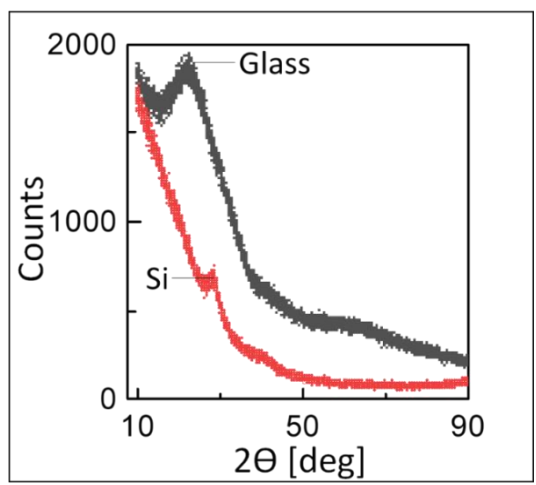

**Figure S1| X-ray diffraction of amorphous molybdenum silicide.** XRD profile of a 15-nm thick $\alpha$Mo$_{81}$Si$_{19}$ film on glass (black) and silicon (red) substrates. Peak of the silicon substrate and low-angle broadening due to the glass substrates are highlighted. The absence of any other significant peaks indicates on the amorphous nature of the films.

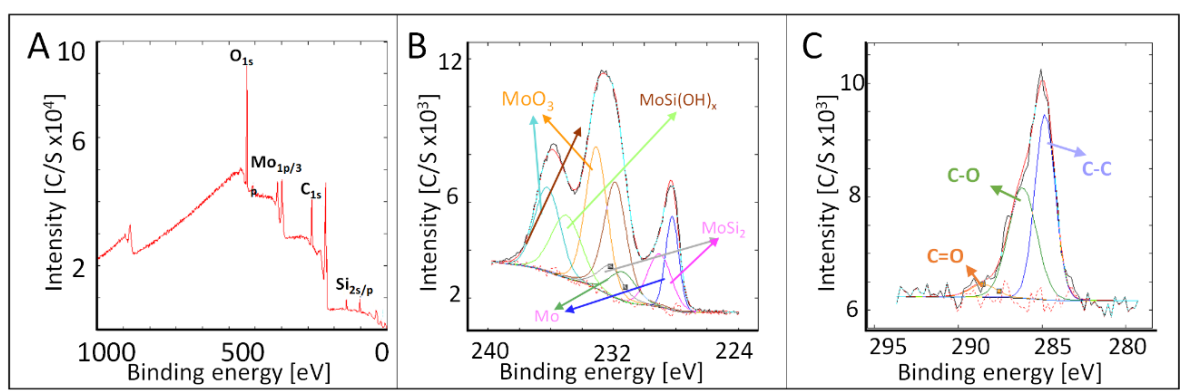

**Figure S2| X-ray photoelectron spectroscopy of molybdenum silicide.** (**A**) Large-scale XPS profile of a Mo$_{1-x}$Si$_x$ film on a silicon substrate shows the film compounds. Here, the Mo is 13.7% and the Si is 7.3% of the complete atomic composition in (A). (**B**) Closer look at the molybdenum binding energies shows that the atomic ratio of the Mo is divided as follows: 14% metal Mo, 37% MoO$_3$, 15% MoSi$_2$, leading to 81:19 Mo:Si composition ratio. (**C**) Closer look at the carbon binding energy that was used for calibration of the stoichiometry.

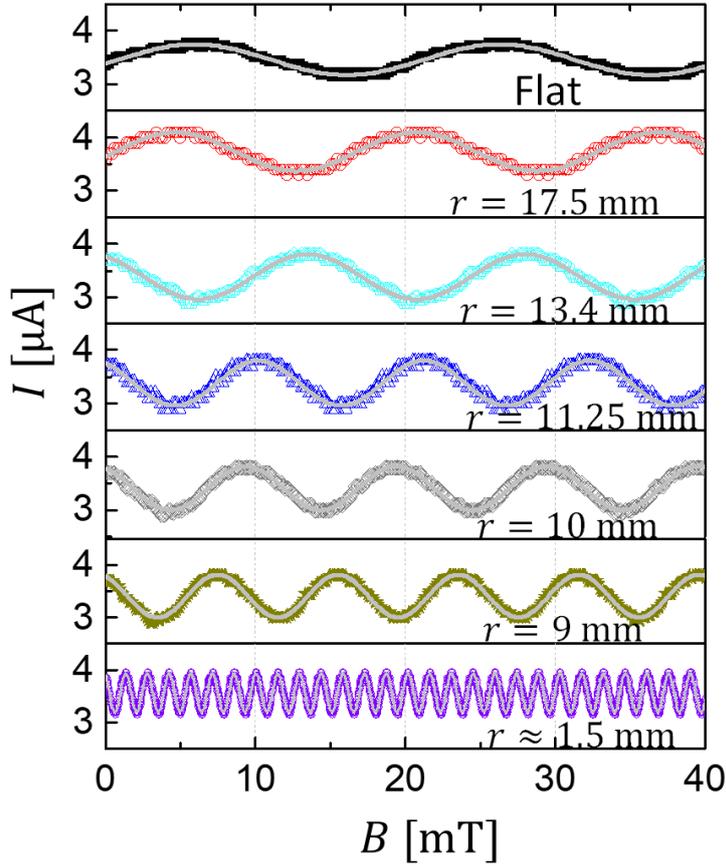

**Figure S3| Effects of bending on interference periodicity in flexible SQUIDs for a large set of curvatures.** Interference pattern of SQUIDs with varying radii of curvature (marked in the figure). Gray lines are the best fits for a sine function. The fitting parameters are given in Table S1.

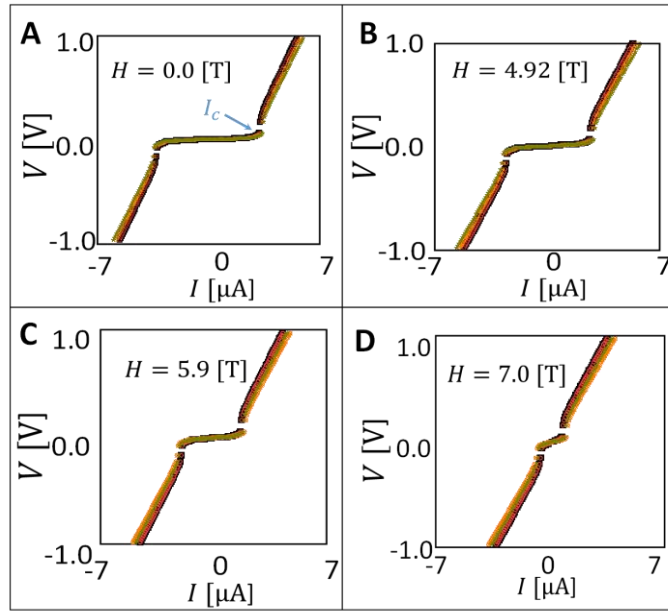

**Figure S4| Current-voltage graphs for different curvatures of an $\alpha Mo_{81}Si_{19}$ nanowire under different magnetic-field values.** *I-V* curves measured for a $100\times2000\times15$ nm$^3$ for various bending conditions (measured in parallel to the SQUID interference measurements in Figures 2 and S3) under (**A**) 0, (**B**) 4.92, (**C**) 5.9 and (**D**) 7 T magnetic fields. Note that the flat (most curved) sample had the highest (smallest) normal resistance (slope at the normal area) for low *H*, but the order was changed for the high magnetic fields.

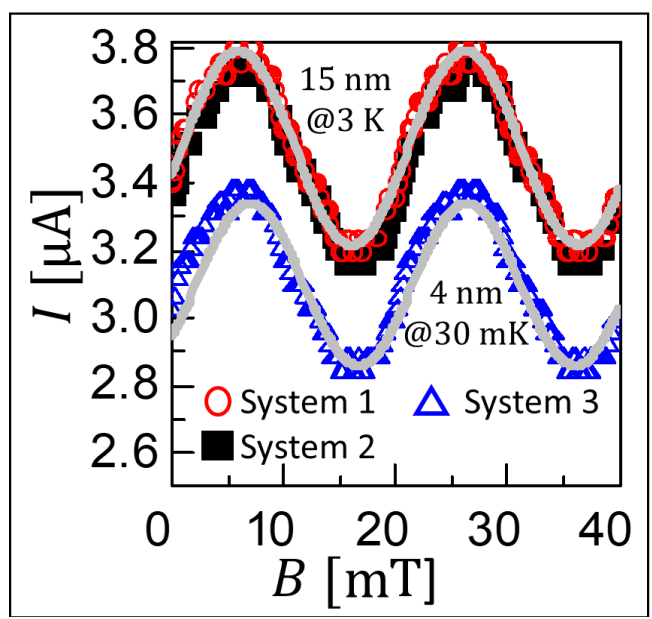

**Figure S5| Comparison between measurements in different testing systems.** Interference pattern of a 15-nm thick SQUID that was characterized at 3 K both in a commercial DynaCool© system (red circles) as well as in an in-house built system (black squares). Blue triangles are an interference pattern of a 4-nm thick SQUID with a similar geometry that was measured at 30 mK with a commercial BlueFors© system, showing the applicability of the device at a broad range of temperatures and thicknesses. The identical profiles of System 1 and 2 as well as the similar periodicity of the flat device with a similar geometry (albeit with different thicknesses) eliminate possible artefacts that might arise due to intrinsic effects of the testing system.

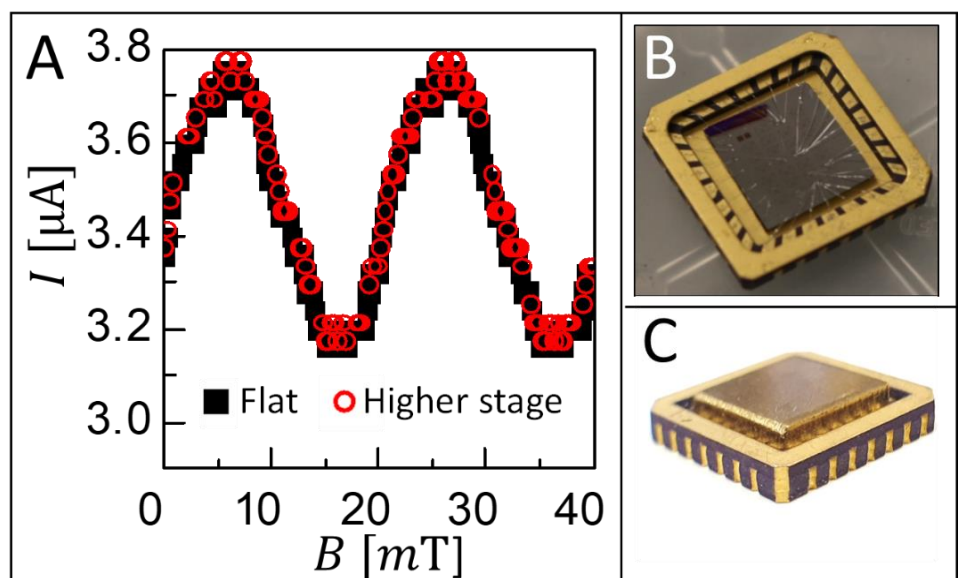

**Figure S6| Effects of sample elevation and sample lifting stage on a SQUID.** (**A**) Interference pattern of a flat SQUID with no curvature and no elevation (black squares) along with the interference pattern of the same device that is elevated on a lifting stage (yet, no curvature) from the same material that was used to curve the SQUID in this work (red circles). Optical photo of the same SQUID at a flat condition (**B**) without and (**C**) with a lifting stage. The identical interference profile eliminates existing of artefacts that arise from the usage of a lifting or curving stage, *e.g.*, due to magnetic properties of the holder of inhomogeneity of the magnetic field in the testing system.

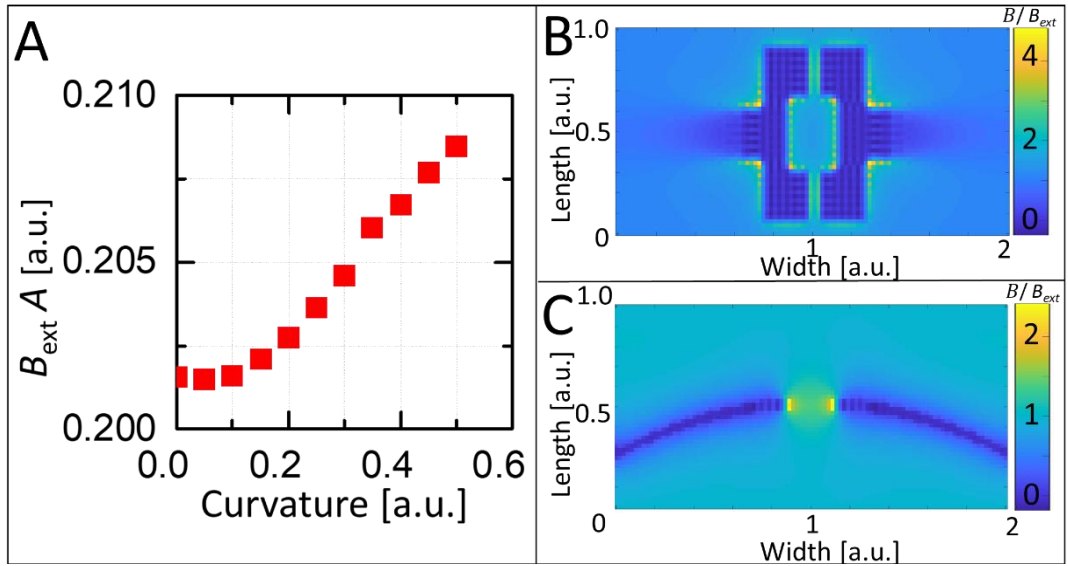

**Figure S7| Contribution of lensing effects on magnetic-field enhancement.** (**A**) Calculated maximal magnetic-field augmentation due to lensing for various curvatures. (**B**) top and (**C**) side view of the simulated magnetic-field enhancement in a device with a substantial curvature (orders of magnitude larger curvature than in our experiments), showing enhancement that is much lower than the observations. Simulations follow Reference (*31*).

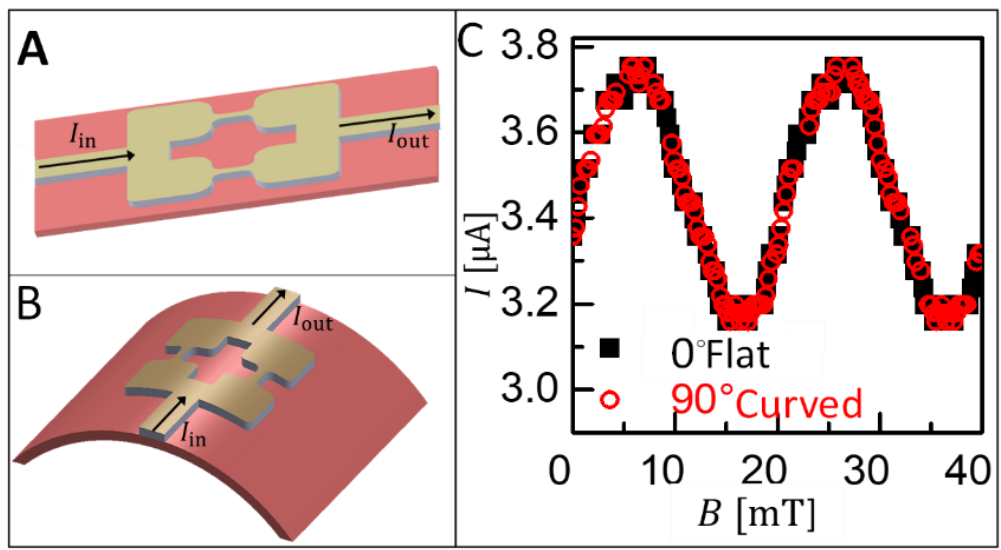

**Figure S8| Comparison between a flat SQUID and a curved SQUID with weak links perpendicular to the curved axis.** (**A**) Schematics of a device that is curved with the weak links perpendicular to the bending axes the flat device as well as of a (**B**) flat device. The device in (A) is rotated in 90° with respect to the device position in all other measurements in this work (see Figure 1A for comparison). (**C**) Interference patterns of a flat SQUID and of the same device that is flexed at $r$=11.2 mm with its weak links perpendicular to the bending direction (red). The two measurements show an identical profile, indicating that the interference pattern changes only if the weak links are parallel and not perpendicular to the strain direction. These data also eliminate various possible experimental artefacts, such as magnetic-field enhancement or inhomogeneity due to the holder, magnetic-field lensing *etc*. (though these results are in general qualitative agreement with Figure S7D).

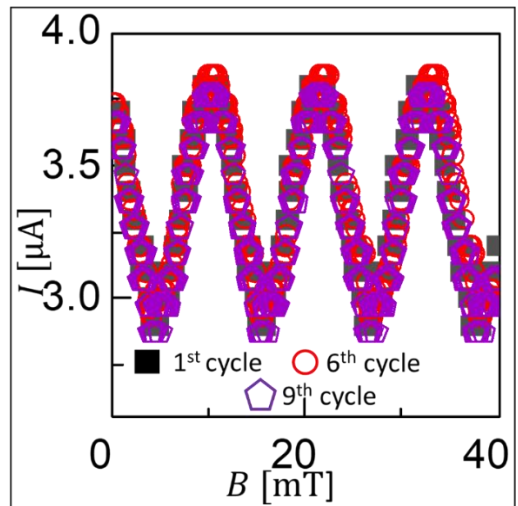

**Figure S9| Repeatability of SQUID behavior under variable flexure conditions and removal-attachment cycles.** Interference pattern of an $\alpha$Mo$_{81}$Si$_{19}$ SQUID placed on a sample holder with $r$=11.2 mm after being placed on the holder for the first time (full black circles) as well as after being removed and attached to the holder by using the adhesion nature of the substrate tap for six times (empty red circle) and for nine times (empty purple pentagons). Note that in between each cycle, the samples were transferred at least two additional times to other holders, so that more than twenty more removal-attachment cycles were done in between. The similarity of these interference pattern demonstrates the robust behavior of the flexible amorphous superconducting materials and devices under varying flexure conditions. Moreover, the data reproducibility helps eliminate the hypothesis that the difference in periodicity for different curvatures stems from *e.g.* changes in the device location in the measurement system due to experimental errors.

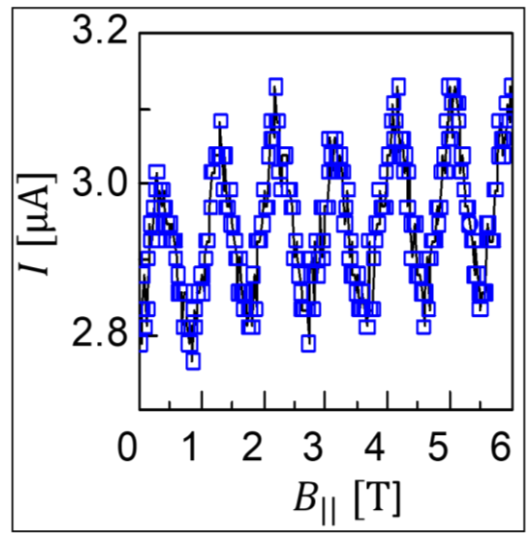

**Figure S10| Amorphous molybdenum silicide SQUID operation under high magnetic fields.** Interference pattern of a 4-nm thick $\alpha$Mo$_{81}$Si$_{19}$ SQUID under high parallel magnetic fields at 30 mK. Here, the interference pattern is induced due to 1.2° misalignment of the parallel field (calculated from the interference pattern due to out-of-plane magnetic field in Figure S5). Note that the 6 T limit was set by the testing system, while our measurements suggest operation even at fields higher than 10 T. High magnetic field operation is advantageous for applications, such as magnetic-resonance imaging (MRI).

| $r$ [mm] | $\kappa$ [mm$^{-1}$] | $B_0$ [mT] | $\delta I_c$ [µA] | $\chi^2$ |
|---|---|---|---|---|
| $\infty$ | 0 | 20.42 | 0.29 | 0.0006 |
| 17.50 | 0.06 | 16.06 | 0.36 | 0.0018 |
| 13.33 | 0.08 | 14.52 | 0.42 | 0.0011 |
| 11.25 | 0.09 | 11.06 | 0.43 | 0.0020 |
| 10.00 | 0.10 | 10.04 | 0.41 | 0.0014 |
| 9.17 | 0.11 | 8.02 | 0.40 | 0.0013 |
| 1.5 | 0.66 | 1.44 | 0.36 | 0.0014 |

**Table S1| SQUID parameters for various flexing curvatures.** Periodicity ($B_0$) and amplitude ($\delta I_c$) of the interference pattern as extracted from the best fits to a sine wave of the data in Figure S3. Statistical errors are much lower than the nominal experimental errors. The statistical parameter $\chi^2$ is introduced for each best-fit analysis.

| SQUID | $r$ [mm] | $\kappa$ [mm$^{-1}$] | $B_0$ [mT] |
|---|---|---|---|
| Large $\alpha$Mo$_{81}$Si$_{19}$ | $\infty$ | 0 | 12.20 |
| | 17.50 | 0.06 | 9.20 |
| | 13.33 | 0.08 | 8.15 |
| | 11.25 | 0.09 | 6.50 |
| | 10.00 | 0.10 | 5.80 |
| | 9.17 | 0.11 | 4.60 |
| Square $\alpha$W$_{60}$Si$_{40}$ | $\infty$ | 0 | 19.88 |
| | 11.25 | 0.09 | 8.4 |

**Table S2| SQUID parameters of different device material and geometry for various flexing curvatures.** Periodicity ($B_0$) of the interference pattern for a large SQUID (Figure 1D) of $\alpha$Mo$_{81}$Si$_{19}$ as well as for a square SQUID of $\alpha$W$_{60}$Si$_{40}$ (insert in Figure 2B) for various curvatures. These data are plotted in Figure 2B.